\documentclass[aps,pra,twocolumn,floatfix]{revtex4}
\usepackage{amsmath}
\usepackage{graphicx}
\usepackage{color}
\usepackage{bm}
\usepackage{bigints}
\usepackage[colorlinks]{hyperref}
	\hypersetup{citecolor=red,urlcolor=blue}
\usepackage{blkarray}
\usepackage{tikz}
\usetikzlibrary{shapes,arrows,automata}
\usetikzlibrary{patterns}
\usetikzlibrary{intersections}

\newcommand{\be}{\begin{eqnarray}}
\newcommand{\ee}{\end{eqnarray}}

\newcommand{\JPB}{\emph {J. Phys. B: At. Mol. Opt. Phys.}}

\newcommand{\PRA}{\emph{Phys. Rev. A}}
\newcommand{\PRL}{\emph{Phys. Rev. Lett.}}

\newcommand{\OEX}{\emph{Opt. Expr.}}
\newcommand{\beq}{\begin{equation}}
\newcommand{\eeq}{\end{equation}}
\newcommand{\beqa}{\begin{eqnarray}}
\newcommand{\eeqa}{\end{eqnarray}}
\newcommand{\nn}{\nonumber}

\newcommand\hlight[1]{}

\newcommand\rhlight[1]{}

\newcommand\ghlight[1]{}

\begin{document}

\title{Electron rotational asymmetry in strong-field photodetachment from F$^-$ by circularly-polarized laser pulses}

\author{G. S. J. Armstrong, D. D. A. Clarke, A. C. Brown and H. W. van der Hart}
\affiliation{Centre for Theoretical Atomic, Molecular and Optical Physics, School of Mathematics and Physics, Queen's University Belfast, University Road, Belfast, BT7 1NN, Northern Ireland
}

\date{\today}
\begin{abstract}
We use the $R$-matrix with time-dependence method to study detachment from F$^-$ in circularly-polarized laser fields of infrared wavelength. By decomposing the photoelectron momentum distribution into separate contributions from detached $2p_1$ and $2p_{-1}$ electrons, we demonstrate that the detachment yield is distributed asymmetrically with respect to these initial orbitals. We observe the well-known preference for strong-field detachment of electrons that are initially counter-rotating relative to the field, and  calculate the variation in this preference as a function of photoelectron energy. The wavelengths used in this work provide natural grounds for comparison between our calculations and the predictions of analytical approaches tailored for the strong-field regime. In particular, we compare the ratio of counter-rotating electrons to corotating electrons as a function of photoelectron energy. We carry out this comparison at two wavelengths, and observe good qualitative agreement between the analytical predictions and our numerical results.
\end{abstract}


\maketitle
\section{Introduction}

Atomic photoemission characteristics induced by circularly polarized fields naturally depart strongly from those of their linearly-polarized counterparts. One distinct feature of circular fields is the possibility of ejecting electrons that rotate with or against the laser polarization. Depending on the polarization direction, it is possible for a particular sign of the magnetic quantum number to be favoured over another, inducing an asymmetric yield distribution among the  accessible magnetic sublevels. This phenomenon, which we term `electron rotational asymmetry', has raised interest in the study of residual-ion ring currents induced by circularly-polarized fields \cite{barth2007,barth2011,herath2012,kaushal2013,kaushal2015,kaushal2018a,kaushal2018b,eckart2018}, as well as in the production of spin-polarized electron bunches \cite{barth2013,hartung2016,liu2017,trabert2018}. This asymmetry has also proved relevant for attoclock measurements using noble gases, particularly in relation to deflection angles of ejected electrons originating from different orbitals. Recent calculations \cite{kaushal2015} demonstrated a noticeable disparity not only in yields, but also in photoelectron offset angles pertaining to $p_1$ and $p_{-1}$ electrons. The calculated angular shifts were deemed large enough to be detected in experiment, thus promoting the attoclock setup as a possible scheme for detecting ring currents.



Experimental techniques for laser polarization control are now well established \cite{brixner2001,brixner2004,misawa2016}, and polarization-tuneable pulses with tailored spectral amplitude and phase are becoming increasingly common \cite{kerbstadt2017a,kerbstadt2017b}. Additionally, laser facilities suitable for high-order harmonic generation (HHG) provide an attractive source of ultrashort pulses of this nature. A classic application of such pulses is the use of bichromatic co- or counter-rotating circularly-polarized laser fields to generate elliptically-polarized attosecond pulse trains \cite{milosevic2000,fleischer2014,milosevic2015,lukas2015,kfir2015}. Harmonic emission from atoms with $p$-electron ground states has recently proved to be an effective and efficient source for such pulse trains \cite{lukas2015,dorney2017}.

With experimental developments gathering pace, theoretical interest in problems using pulses of circular and elliptical polarization is rising. However, accurate treatment of the dynamics induced by circularly-polarized fields poses significant challenges for theoretical methods. The necessary inclusion of magnetic sublevels quickly scales calculations to a daunting size. Therefore, only a limited number of {\em ab initio} methods have been developed for electron dynamics in arbitrarily polarized light fields. The response of the H atom exposed to circularly polarized pulses has been studied using a variety of methods \cite{askeland2011,bauer2014,wu2016}. Recently, a two-electron approach \cite{djiokap2014,djiokap2016,djiokap2017} has investigated both single- and double-ionization dynamics induced by pulses of circular and elliptical polarization. A similar computational technique has been used to calculate emission characteristics in multiphoton double ionization of H$_2$ in circularly-polarized pulses \cite{pindzola2017}. Strong-field single-ionization of molecular ions in bichromatic, circularly-polarized pulses has also been studied \cite{yuan2016,yuan2017,yuan2018}, with a focus on HHG.

Such calculations have provided detailed insights into the dynamics of one- and two-electron systems in arbitrarily-polarized fields. However, time-dependent simulations for multielectron atoms  exposed to such fields have thus far relied mainly on the single-active-electron approximation (SAE) \cite{kjeldsen2007,abu2011,ivanov2013,ivanov2014,mancusco2015,ilchen2017}. This is in spite of an extensive body of experimental and theoretical evidence challenging the SAE perspective \cite{weber2000,drescher2002,uiberacker2007,leone2014}, suggesting that a correlated, multielectron response to the laser field can be decisive for an authentic characterization of the dynamics.


The {\em ab initio} and fully nonperturbative treatment of laser interactions with multielectron atoms is a demanding theoretical and computational task. Indeed, both the electronic structure of the irradiated target, as well as the strong-field ionization dynamics, must be reliably captured. To date, one of the few methodologies that achieves this is the $R$-matrix with time dependence (RMT) theory \cite{moore2011}. The RMT method has been used to analyze strong-field dynamics of a variety of atoms and ions \cite{hassouneh2014,brown2016,clarke2017}, and has recently been extended to handle light fields of arbitrary polarization \cite{clarke2018}. A recent RMT study \cite{hassouneh2015}, investigated above-threshold detachment and electron rescattering in the F$^-$ negative ion, driven by long-wavelength (1300 nm and 1800 nm), ultrashort laser pulses. Comparison with a Keldysh-type approach \cite{gribakin1997} highlighted the role of short-range multielectron correlations, and underlined the need for many-body, quantum-dynamic simulations to obtain reliable photoelectron spectra.




More generally, electron detachment from negative ions is of considerable interest, since their structure, and laser-driven dynamics, are significantly influenced by multielectron correlations. Various techniques have been used to treat photodetachment from negative ions \cite{gribakin1997,beiser2004,bergues2005,bivona2006,krajewski2006,becker2010,shearer2013,hassouneh2015,milosevic2016}, where electron repulsion is handled approximately.  Negative ions are particularly amenable to investigation by methods based on the strong-field approximation (SFA) \cite{becker2010}, since their neutral core dictates that the detachment dynamics is influenced by short-range interactions alone. Investigations on the response of negative ions to circularly-polarized fields are becoming more prevalent \cite{bergues2005,milosevic2016,pisanty2017}. Additionally, recent SFA calculations of photoelectron momentum distributions for F$^-$, subject to orthogonal laser fields, have shown good agreement with those obtained by solving the time-dependent Schr\"{o}dinger equation within the SAE approximation \cite{chen2018}.

For general atoms and ions in strong fields of circular polarization, the dependence of the ionization or detachment rate on the bound electron magnetic quantum number may be given approximately by a number of analytical methods. Among the most widely-used of these methods is Perelomov, Popov and Terent'ev (PPT) theory \cite{perelomov1967,perelomov1967b,perelomov1967c}. Recently, PPT theory was used to investigate strong-field ionization of Kr $4p$ electrons in circular fields \cite{barth2011}, and predicted that throughout a wide range of laser frequencies and intensities, the vast majority of photoelectrons initially rotated against the laser polarization. Analogous expressions may also be obtained using the analytical $R$-matrix (ARM) technique \cite{kaushal2013,kaushal2015,kaushal2018a,kaushal2018b}, which can also account for Coulomb interactions when necessary. A notable recent study of ring currents in Ar demonstrated strong agreement between measured photoelectron spectra, ARM predictions, and those obtained from SAE simulations \cite{eckart2018}. Here it was predicted that counter-rotating electrons dominate the ionization signal. However, it was also apparent that high-energy emission could arise from either co- or counter-rotating electrons.

In this paper, we employ the RMT method \cite{clarke2018} to study electron detachment from F$^-$ by circularly-polarized laser fields. We study the propensity for detachment of valence $2p$ electrons of different magnetic quantum numbers (i.e.\;electrons co- and counter-rotating relative to the laser polarization) as a function of laser wavelength. This propensity is well-known for both one-photon and multi-photon ionization, but the intermediate few-photon regime has not been systematically studied. Here we attempt to gain a better understanding of the transient preference for co-rotating and counter-rotating electrons between the one-photon and multiphoton regimes. We also investigate how this preference varies with the number of above-threshold photon absorptions. We compare our {\it ab initio} predictions with those of the ARM and PPT analytical methods, enabling a rigorous assessment of their qualitative reliability over a range of laser wavelengths.


\section{Theoretical method}

The theoretical foundation of the RMT method for general multielectron atoms in arbitrarily-polarized laser fields is described in Ref. \cite{clarke2018}. The multielectron time-dependent Schr\"{o}dinger equation is solved in the electric dipole and non-relativistic approximations. Fundamentally, position space is divided into two distinct regions, according to radial distance from the nucleus. An inner region is confined to small distances, and encapsulates the target nucleus. This region contains a truly many-body wave function, that accounts for both electron exchange and electron-electron correlation. An outer region extends to relatively large radial extent, and contains a single, ionized electron that is subject to the long-range, multipole potential of the residual system, as well as the laser field.


RMT uses a hybrid numerical scheme, comprised of basis-set and finite-difference techniques. In the inner region, the time-dependent, $(N+1)$-electron wave function is represented by an expansion in $R$-matrix basis functions, with time-dependent expansion coefficients. These basis functions are generated from the $N$-electron wave functions of the residual ion states as well as from a complete set of one-electron continuum functions describing the motion of the ejected electron. The outer-region wave function is constructed using residual-ion wave functions and radial wave functions of the ejected electron in each channel. A finite-difference discretization scheme is used to represent the ejected-electron wavefunction.

In both regions the laser-atom interaction is treated in the length gauge, a choice merited by previous calculations \cite{hutchinson2010}. A velocity-gauge interaction tends to emphasize short-range excitations close to the nucleus.  Accurate capture of such excitations requires a highly-detailed description of the target atomic structure. The length-gauge interaction reduces the effect of such excitations, thereby allowing a less elaborate atomic structure.



\section{Calculation parameters}
\label{calcp}

Our treatment of the F$^{-}$ ionic structure is described in previous work \cite{hassouneh2015,clarke2018}, and is based on earlier $R$-matrix Floquet calculations for this system \cite{vdh1996,vdh2000}. Within the inner region, the neutral F atom is described using a set of Hartree-Fock $1s$, $2s$ and $2p$ orbitals for the F ground state, using the data of Clementi and Roetti \cite{clemroe}. To improve our description of the residual neutral, we include $\overline{3s}$, $\overline{3p}$ and $\overline{3d}$ pseudo-orbitals \cite{niang}. Inclusion of these pseudo-orbitals enables a more accurate determination of the $1s^2 2s^2 2p^5$ $^2 P ^o$ F ground-state wave function, by performing a configuration-interaction calculation that includes the $1s^2 2s^2 2p^5$, $1s^2 2s 2p^5 3s$, $1s^2 2s^2 2p^4 3p$, $1s^2 2s^2 2p^3 3p^2$ and $1s^2 2s^2 2p^3 3d^2$ configurations. Our atomic structure calculations yield a binding energy of 3.42 eV for the initial $^1S^e$ F$^-$ ground state, which agrees well with the experimental value of 3.401 eV \cite{blondel2001}.

To test the sensitivity of the results to the atomic structure description of the residual neutral, we also employ an additional, simpler model of the F atom. This model includes only the $1s^2 2s^2 2p^5$ configuration, using the Hartree-Fock $1s$, $2s$ and $2p$ orbitals \cite{clemroe}. The binding energy obtained in this calculation is artificially shifted to match the value of 3.42 eV obtained in the model that includes pseudo-orbitals.

The radial extent of the inner region is 50 a.u., which suffices for effective confinement of the orbitals of the F$^-$ ion. The inner-region continuum functions are generated using a set of 60 $B$-splines of order 13 for each available orbital angular momentum of the outgoing electron. We retain all admissible $1s^22s^22p^5\epsilon l$ channels up to a maximum total angular momentum $L=L_{\rm max}$, as well as the required set of magnetic sublevels.
Calculations for circular polarization are most efficiently performed using pulses polarized in the $xy$ plane, as this geometry halves the number of dipole-accessible symmetries \cite{clarke2018}. The two-photon calculations at 532 nm required $L_{\rm max}=12$ for satisfactory convergence. The calculations at wavelengths of {\mbox{800 nm}} and {\mbox{1064 nm}} used $L_{\rm max} = 29$, a setting that yielded 900 $LM_LS\pi$ symmetries, and 1364 channels. At 1560 nm, $L_{\rm max} = 49$ was required, resulting in 2500 $LM_LS\pi$ symmetries, and 3774 channels.



In the outer region, the radial motion of the ejected-electron is treated using a one-dimensional radial grid, with uniform mesh size $\delta r = 0.08$ a.u.. We adopt a fifth-order finite-difference scheme, which ensures a high degree of accuracy in describing the spatial properties of the ionized-electron wave packet. The outer-region grid extends to large radial distances, enabling the asymptotic characteristics of the ionized-electron wave function (and hence, the photoelectron energy and momentum distributions) to be reliably ascertained. The calculations for a 532-nm laser wavelength use a radial grid with a maximum extent of 3400 a.u. . For the 800-nm and 1064-nm wavelengths we use a maximum radial extent of 2880 a.u., and at 1560 nm a maximum extent of 4400 a.u. is used. 

The main observable of interest in this work is the photoelectron momentum distribution, a quantity that is particularly sensitive to the details of the time propagation. A high-accuracy time propagation of the wave function is secured by using an 8th-order Arnoldi propagator \cite{moore2011}, with a timestep $\delta t = 0.01$ a.u.. Accurate momentum distributions may then be obtained by ensuring that the wave function is propagated for a significant length of time after the pulse has terminated. At 532 nm, 800 nm and 1064 nm, the total propagation time is 2000 atomic units, which represents 12 cycles and 8 cycles following termination of the respective pulses. At 1560 nm, the total propagation time is 2800 atomic units, equivalent to 13 field cycles, with the pulse terminating after 8 cycles. 

Following the time propagation, we obtain the radial part of the ejected-electron wave function in each channel. However, the full outer-region wave function is represented by an expansion in channel functions \cite{lysaght2009,clarke2018}, which couple the orbital and spin angular momenta of the outgoing electron to those of the residual F neutral. 
Therefore, in each channel, the radial and angular dependence of the outgoing electron wave function is obtained after decoupling its angular momenta from those of the residual system \cite{vdh2008}.
Once acquired, the wave function is transformed, for $r > 50$ a.u., into the momentum representation by means of a standard Fourier transform. Analysis of the channel wave functions shows that the lowest-energy wave packets have reached radial distances of at least $r > 50$ a.u.\;by the final propagation time, and possess continuum character, indicating that the entire photoelectron wave function is faithfully transformed to momentum space.

 
 
The F$^-$ target interacts with a laser field that is treated classically within the electric dipole approximation. Since the laser interaction is described in the length gauge, we adopt a carrier-envelope form for the electric field, circularly polarized in the $xy$ plane. The right-hand circularly-polarized fields used in this work take the form
\beq
{\cal E}(t) = \frac{{\cal E}_0}{\sqrt{2}} \sin^2\left(\frac{\omega t}{2N_c}\right) 
\left[
\cos\omega t \;\hat{\bf x} + \sin\omega t \;\hat{\bf y}
\right],
\label{efield}
\eeq
where ${\cal E}_0$ is the peak electric field strength, $\omega$ is the laser frequency, and $N_c$ is the number of laser cycles. The peak intensity $I_0$ is related to the electric field strength using $I_0 = c{\cal E}_0^2/4\pi$, where $c$ is the speed of light in vacuum.

In this work, the 532-nm, 800-nm and 1064-nm laser pulses ramp on over 3 laser cycles, followed by 3 cycles of ramp-off, so that $N_{c} = 6$ in these cases. At 1560 nm, the pulse ramps on and off over 4 cycles, so that $N_{c} = 8$.  Calculations are performed for a range of laser wavelengths, from three-photon detachment ($\lambda=800$ nm) to five-photon ($\lambda=1560$ nm). At 1560 nm, the peak laser intensity is {\mbox{$I_0 = 2\times10^{12}$ W/cm$^2$}}. In all other cases, the peak intensity is {\mbox{$I_0 = 3\times10^{12}$ W/cm$^{2}$}}.

\section{Results and Discussion}

We now discuss application of RMT to few-photon detachment from F$^-$ in a circularly polarized laser field. In such fields, the dependence of the detachment characteristics on the sign of the valence-electron magnetic quantum number $m_l$, is of considerable interest. In both one-photon ionization and ionization from Rydberg states of hydrogen by microwave fields \cite{rzazewski1993,zakrzewski1993}, it is well known that a circularly-polarized field preferentially ionizes corotating electrons. On the other hand, recent studies \cite{herath2012,barth2011,kaushal2013,kaushal2015,kaushal2018a,kaushal2018b,eckart2018} have established that counter-rotating electrons are preferentially ionized in the strong-field regime. 

In the following sections, we determine the ejected-electron rotational asymmetry, resolved in photoelectron momentum, for F$^-$ in circularly-polarized, near-infrared laser pulses. We place particular emphasis on the variation of this asymmetry with laser wavelength. Moreover, we compare the ratio of ionization yields for $p_{\pm 1}$ electrons, suggested by our RMT calculations, to that predicted by PPT and ARM theories, previously employed for strong-field ionization problems.

 \subsection{Selection of electron rotation}

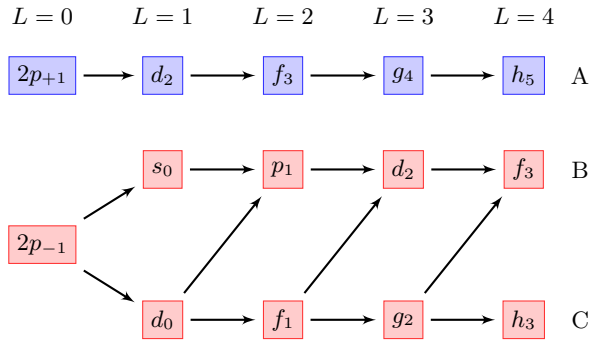
\begin{figure}[t]

 \tikzstyle{square}=[rectangle,thick,minimum size=0.5cm,draw=blue!80,fill=blue!20]
 \tikzstyle{vspecies}=[rectangle, minimum size=0.5cm,draw=blue!80,fill=blue!20]
 \tikzstyle{square}=[rectangle,thick,minimum size=0.5cm,draw=red!80,fill=blue!20]
 \tikzstyle{fspecies}=[rectangle, minimum size=0.5cm,draw=red!80,fill=red!20]

 \begin{tikzpicture}[auto, outer sep=3pt, node distance=1.6cm,>=latex']
 \node [vspecies] (S) {$2p_{+1}$};
 \node [above of  = S, node distance = 0.8cm] (L0) {$L=0$};
 \node [vspecies, right of = S] (d2) {$d_{2}$};
 \node [above of  = d2, node distance = 0.8cm] (L1) {$L=1$};
 \node [vspecies, right of = d2] (f3) {$f_{3}$};
 \node [above of  = f3, node distance = 0.8cm] (L2) {$L=2$};
 \node [vspecies, right of = f3] (g4) {$g_{4}$};
  \node [above of  = g4, node distance = 0.8cm] (L4) {$L=3$};
 \node [vspecies, right of = g4] (h5) {$h_{5}$};
  \node [above of  = h5, node distance = 0.8cm] (L4) {$L=4$};
 \draw [->,thick] (S) --  node {}(d2);
 \draw [->,thick] (d2) --  node {}(f3);
 \draw [->,thick] (f3) --  node {}(g4);
 \draw [->,thick] (g4) --  node {}(h5);
 \node [right of = h5,node distance=0.75cm](A){A};

 \node [fspecies, below of = S,node distance=2.25cm] (Sm) {$2p_{-1}$};
\node [fspecies, below of = d2,node distance=1.25cm] (s0) {$s_{0}$};
 \node [fspecies, below of = d2,node distance=3.25cm] (d0) {$d_{0}$};
 \node [fspecies, right of = s0] (p1) {$p_{1}$};
 \node [fspecies, right of = d0] (f1) {$f_{1}$};
 \node [fspecies, right of = p1] (d2m) {$d_{2}$};
 \node [fspecies, right of = f1] (g2) {$g_{2}$};
 \node [fspecies, right of = d2m] (f3m) {$f_{3}$};
 \node [fspecies, right of = g2] (h3) {$h_{3}$};

 \node [right of = f3m, node distance=0.75cm]{B};
 \node [right of = h3, node distance=0.75cm]{C};

 \draw [->,thick] (Sm) --  node{}(s0);
 \draw [->,thick] (Sm) --  node[below left]{}(d0);
 \draw [->,thick] (s0) --  node{}(p1) ;
 \draw [->,thick] (d0) --  node[below]{}(f1);
 \draw [->,thick] (p1) --  node{}(d2m) ;
 \draw [->,thick] (f1) --  node[below]{}(g2);
 \draw [->,thick] (d2m) --  node{}(f3m) ;
 \draw [->,thick] (g2) --  node[below]{}(h3);

 \draw [->,thick] (d0) --  node [left]{}(p1);
 \draw [->,thick] (f1) --  node [left]{}(d2m);
 \draw [->,thick] (g2) --  node [left]{}(f3m);

 \end{tikzpicture}

 \caption{Lowest-order pathways for detachment of $p_1$ (path A) and $p_{-1}$ (paths B and C) electrons in a right-hand circularly-polarized laser field (Eq.\;\eqref{efield}). Ejected-electron $l$ and $m_l$ values and total orbital angular momentum $L$ are indicated. Detachment of $p_0$ electrons is not considered, as their detachment yield is negligible in the polarization ($xy$) plane.
 }
 \label{paths}
 \end{figure}

To quantify the contributions of initially bound $2p_{\pm1}$ electrons, we decompose the momentum distribution according to the photoelectron magnetic quantum number, $m_l$. To do this, we consider the contribution of specific electron-detachment channels, that we identify using the pathways shown in Fig.\;\ref{paths}. In a right-hand circularly polarized laser field, the selection rule on the single-electron $m_l$ value is $\Delta m_l = 1$. Detachment of a $p_1$ (corotating) electron therefore proceeds via a single pathway, passing through a set of possible final channels indicated in Fig.\;\ref{paths} as set A.
 Detachment of a $p_{-1}$ (counter-rotating) electron proceeds along two possible paths, accessing two sets of final channels, labelled B and C. 
 Symmetry arguments dictate that $p_0$ electrons, aligned perpendicular to the polarization plane, make a negligible contribution to the momentum distribution in this plane \cite{corkum2007}. 
 
 \subsection{Two-photon detachment}

 \begin{figure*}[t]
 $\arraycolsep=1.0pt\def\arraystretch{0.6}
 \begin{array}{ccc}
 {(a) \ \textrm{corotating}} &  {(b) \ \textrm{counter-rotating}} & (c) \ \textrm{angle-integrated spectrum}
 \\
 {\centering\includegraphics[trim=0.0cm 1cm 1cm 1cm,clip=true,width=0.68\columnwidth]{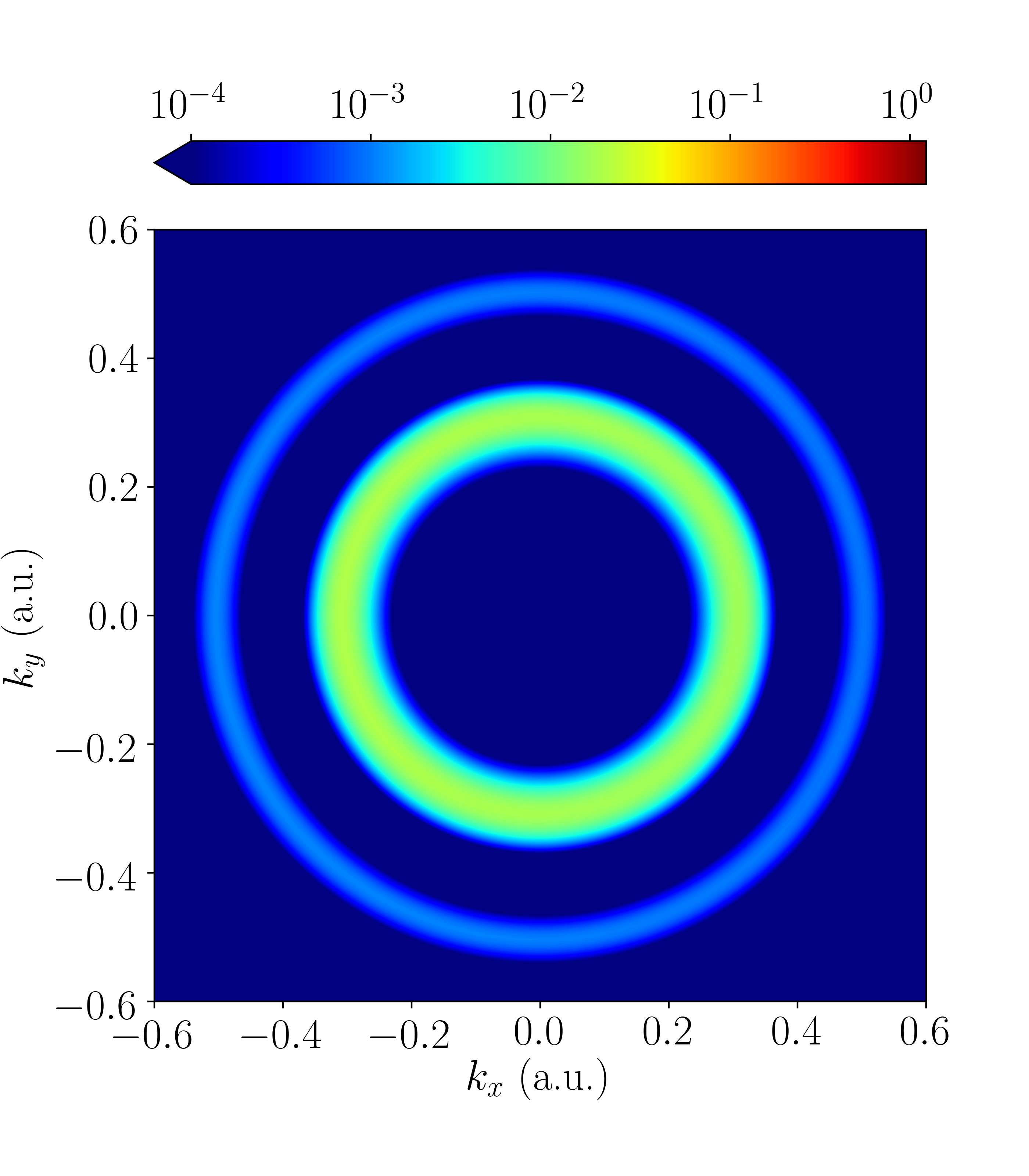}}
 &
 {\centering\includegraphics[trim=0.0cm 1cm 1cm 1cm,clip=true,width=0.68\columnwidth]{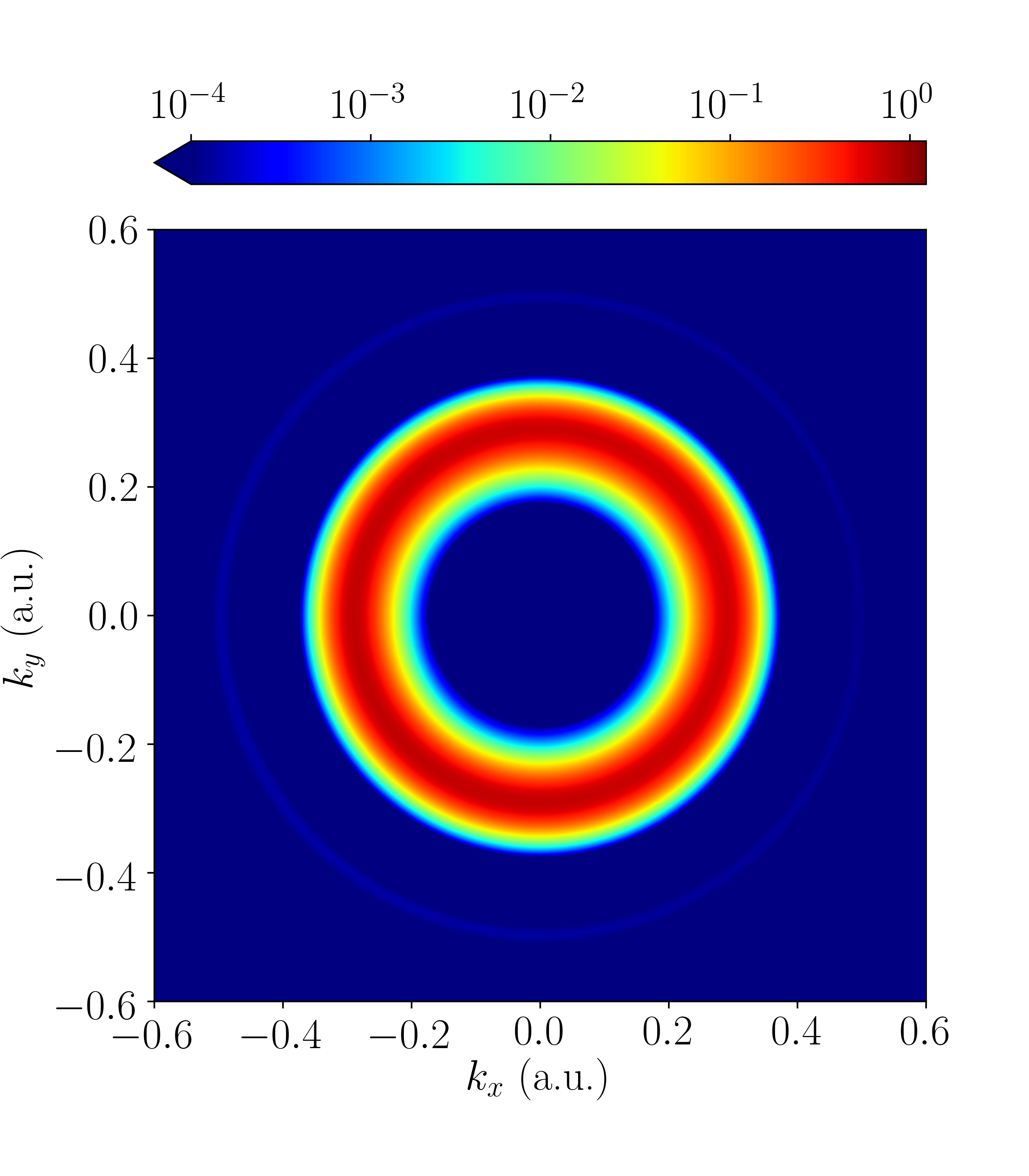}}
 &
 {\centering\includegraphics[width=0.62\columnwidth]{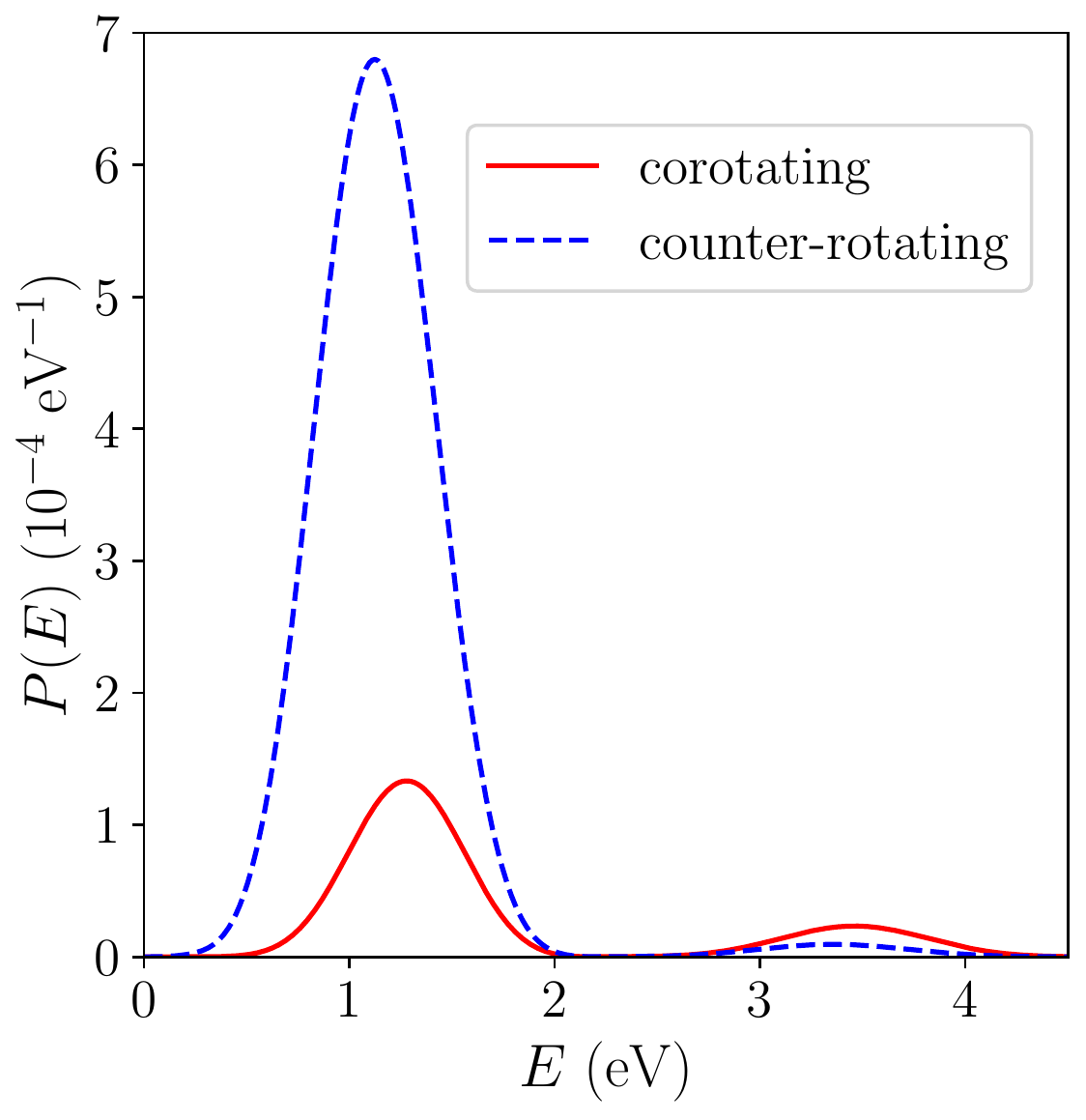}}
 \end{array}$
 \caption{Photoelectron momentum distributions in the polarization plane, following two-photon detachment from F$^-$, initiated by a right-hand circularly polarized laser pulse with a carrier wavelength of 532 nm, a duration of $N_c = 6$ cycles, a peak intensity of $I_0 = 3 \times 10^{12}$ Wcm$^{-2}$.  The angle-integrated distributions (c) (integrated over azimuthal angle $\phi$) demonstrate the differing energy dependence of co- and counter-rotating electrons in the polarization plane. The data presented here can be accessed via Ref.\;\cite{pure}.} 
 \label{2wfig}
 \end{figure*}

Figure \ref{2wfig} presents photoelectron momentum and energy distributions in the polarization plane, for two-photon detachment from F$^-$, induced by a 6-cycle, 532-nm, right-hand circularly polarized pulse. The momentum distributions shown in Figs.\;\ref{2wfig}(a) and (b) are comprised of concentric toroidal features. The innermost toroid arises from two-photon detachment, and the outer toroid from above-threshold detachment (ATD) following three photon absorptions. At this laser intensity ($3\times10^{12}$ W/cm$^2$), higher-order processes are negligible.

Figures\;\ref{2wfig} (a) and (b) contain two main features of interest. Firstly, counter-rotating electrons clearly dominate the total yield, indicating a strong rotational asymmetry. Although striking in its own right, this observation is further distinguished by its stark contrast with the well-established preference for detachment of co-rotating electrons by a single photon. Indeed, such a preference was observed in recent RMT calculations for F$^-$ \cite{clarke2018}, with corotating electrons taking an 84\% share of the single-photon yield. In the two-photon detachment calculation presented here, the change in preference is rather dramatic --- counter-rotating electrons provide almost 80\% of the total detachment yield.

The distributions contain a further salient feature --- at the first above-threshold peak the rotational preference is reversed. Despite the low yield at the first ATD peak, a preference for corotating electrons is evident, and therefore the degree of rotational asymmetry varies strongly with excess energy. The energy-dependent nature of the rotational asymmetry is made clear by calculating the energy distribution of co- and counter-rotating electrons in the polarization plane, by integrating the distributions of Fig.\;\ref{2wfig}(a) and (b) over the azimuthal angle. The respective angle-integrated distributions are shown in Fig.\;\ref{2wfig}(c), and the reversal in preference as energy varies is apparent.

\subsection{Three-photon detachment}

\begin{figure*}[t]
$\arraycolsep=1.0pt\def\arraystretch{0.6}
\begin{array}{ccc}
(a) \ \textrm{corotating} &  (b) \ \textrm{counter-rotating}
&  (c) \ \textrm{angle-integrated spectrum}
\\
{\centering\includegraphics[trim=0.0cm 1cm 1cm 1cm,clip=true,width=0.68\columnwidth]
{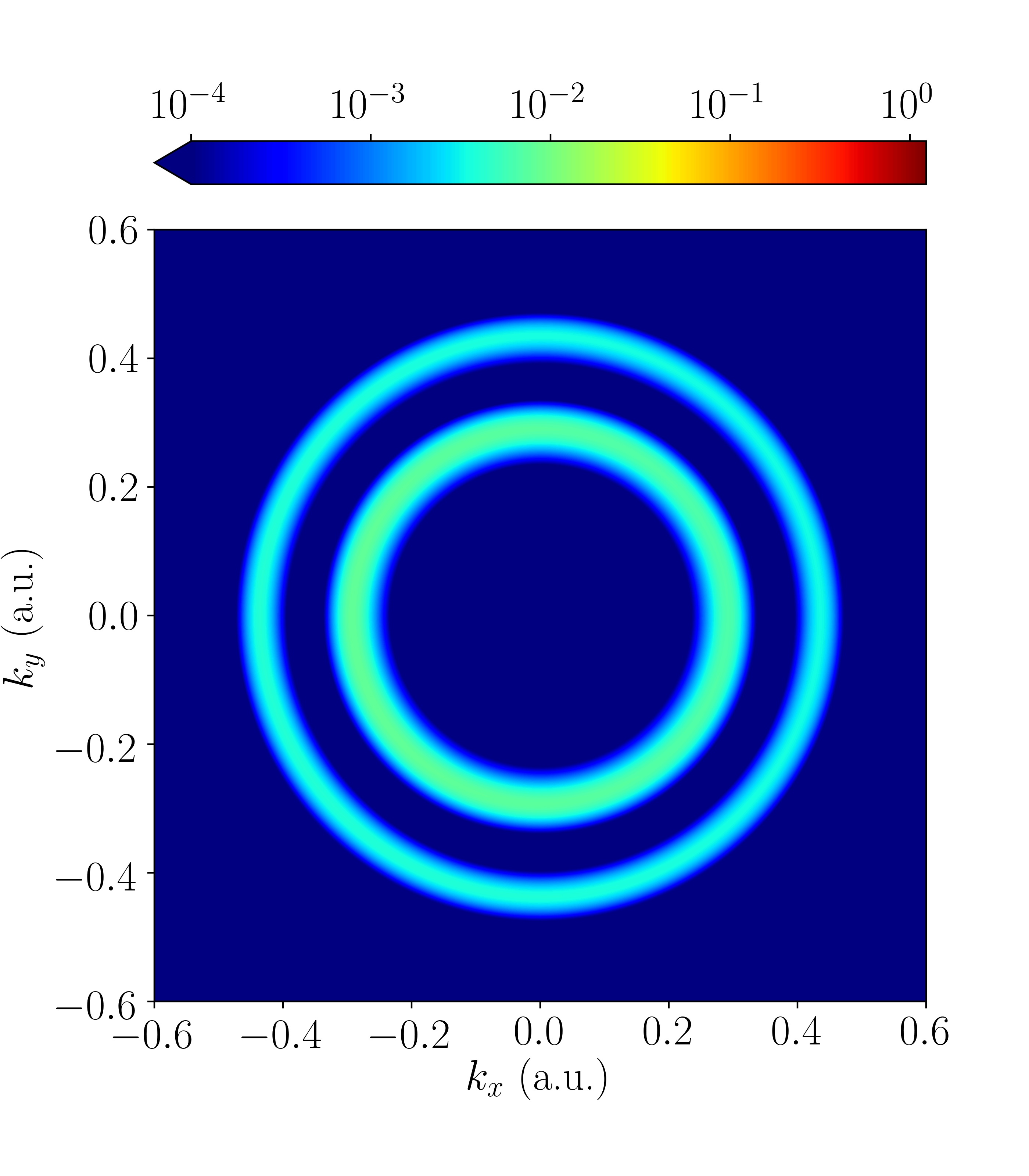}
}
&
{\centering\includegraphics[trim=0.0cm 1cm 1cm 1cm,clip=true,width=0.68\columnwidth]
{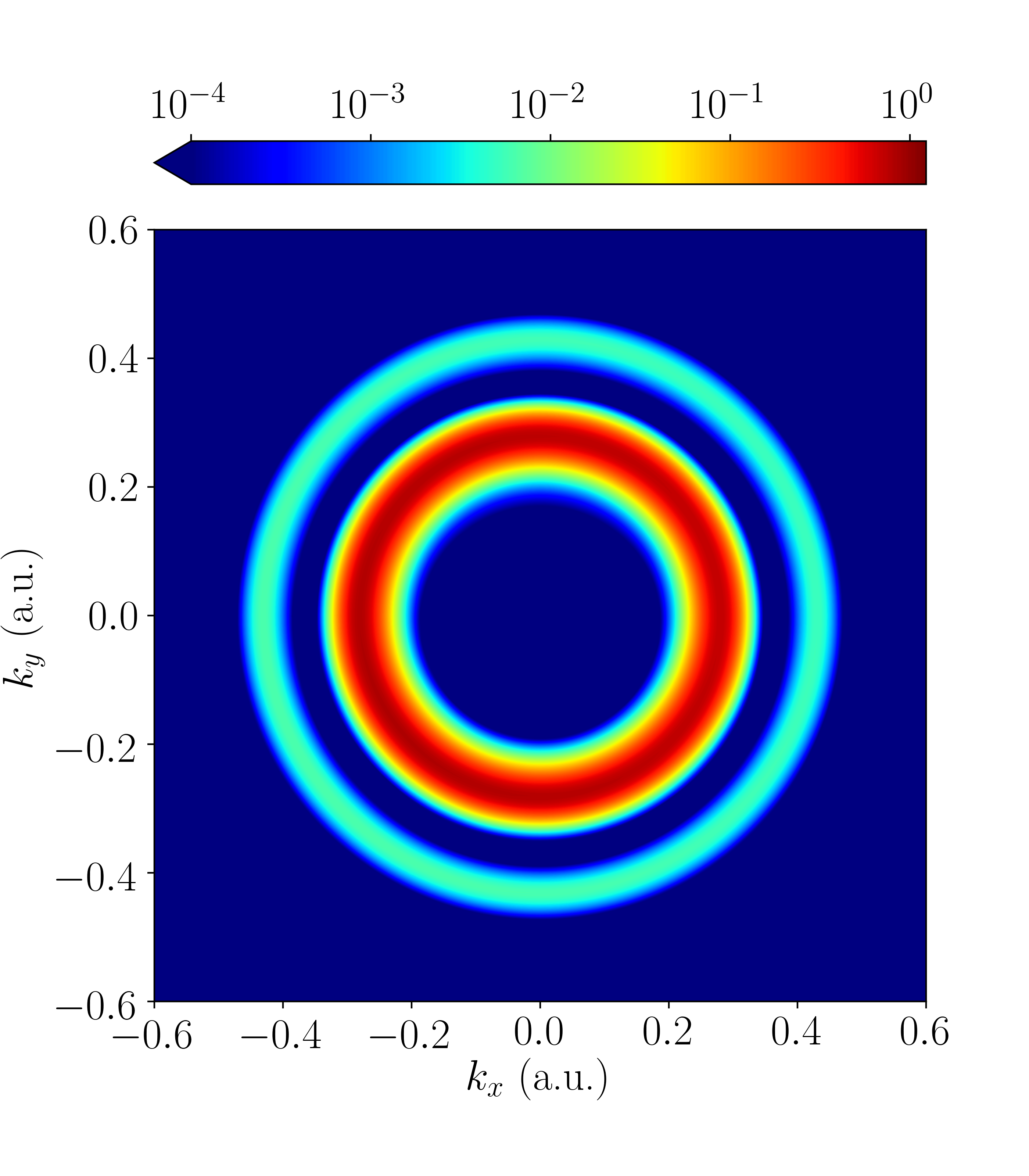}
}
&
\includegraphics[width=0.64\columnwidth]{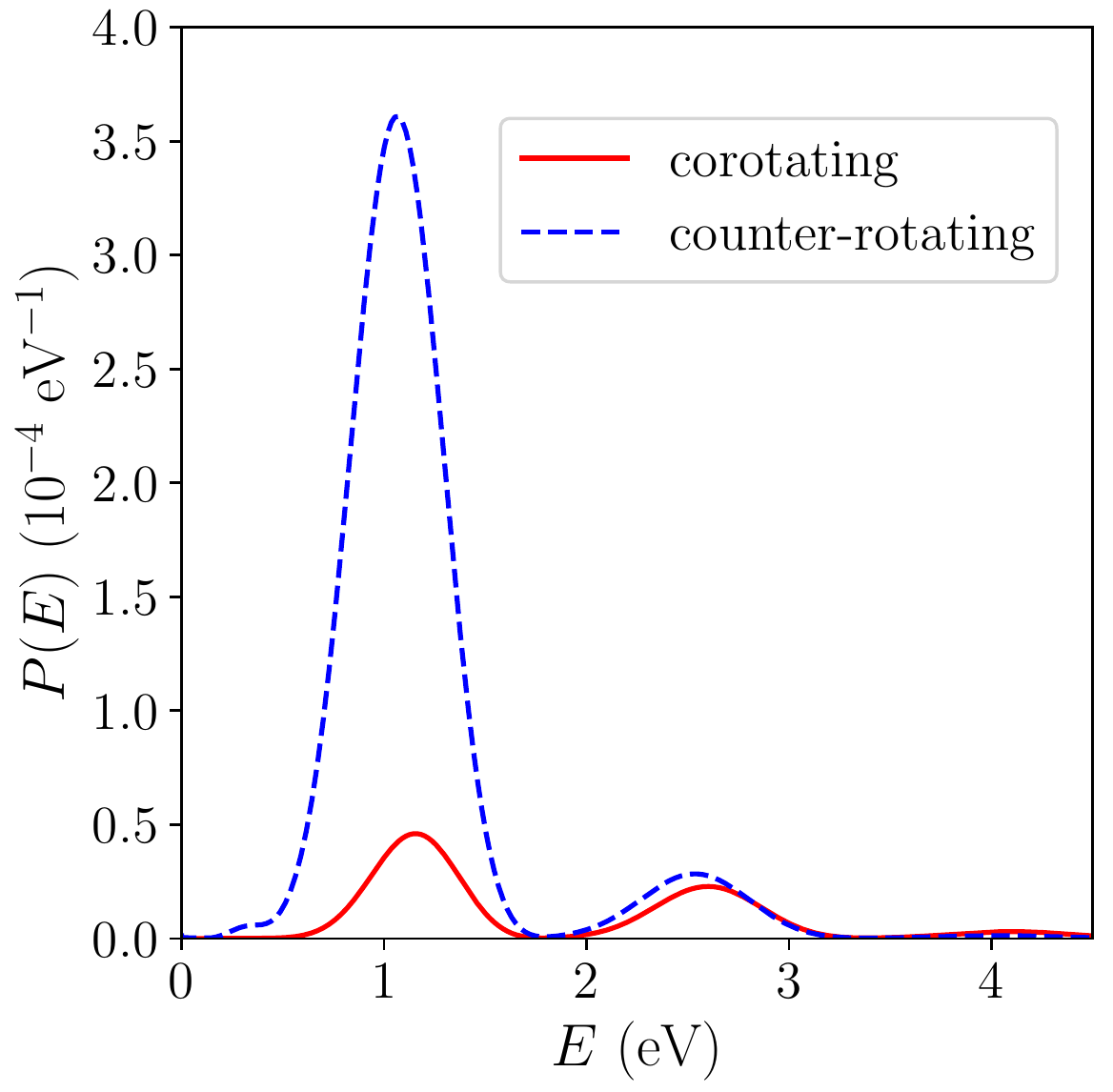}
\end{array}$
\caption{(Color online) Photoelectron momentum distributions in the polarization plane for (a) corotating and (b) counter-rotating electron detachment, following three-photon detachment from F$^-$ by a right-hand circularly polarized laser pulse with a carrier wavelength of 800 nm, a duration of $N_c = 6$ cycles, and a peak intensity of $I_0 = 3 \times 10^{12}$ Wcm$^{-2}$.  The angle-integrated distributions (c) (integrated over azimuthal angle $\phi$) demonstrate the differing energy dependence of co- and counter-rotating electrons. } 
\label{3wfig}
\end{figure*}

Figure \ref{3wfig} presents photoelectron momentum and energy distributions in the polarization plane, for three-photon detachment from F$^-$, induced by a 6-cycle, 800-nm, right-hand circularly polarized pulse. Here, the innermost toroid arises from three-photon detachment, and the outer toroid from above-threshold detachment (ATD) following four photon absorptions. At this choice of laser intensity, higher-order processes are again negligible.

Comparing Figs.\;\ref{3wfig} (a) and (b), two features are striking. Firstly, a strong rotational asymmetry is visible, with counter-rotating electrons clearly dominating the total yield. Again, this contrasts with the well-established preference for single-photon detachment of co-rotating electrons, observed for F$^-$ in previous RMT calculations \cite{clarke2018}. 
For three-photon detachment, counter-rotating electrons now contribute over 80\% of the total yield.


Secondly, the degree of asymmetry once again varies strongly with excess energy. The lowest-energy electrons close to threshold are predominantly counter-rotating, whereas the first ATD peak displays less disparity. In fact, the yields for three-photon and four-photon detachment of co-rotating electrons are comparable, whereas the analogous yields for detachment of counter-rotating electrons differ by more than an order of magnitude. The respective energy distributions of co- and counter-rotating electrons in the polarization plane are shown in Fig. \ref{3wfig}(c). The degree to which counter-rotating electrons dominate the low-energy yield is evident, as is the level of parity obtained between the contributions at the first above-threshold peak.

We note one further feature of the distributions, namely the apparent discrepancy in the widths of the peaks in Fig.\;\ref{3wfig}(a) and (b) respectively. This is simply due to the choice of scale, and in fact the respective peaks have almost identical widths, largely dictated by the laser bandwidth.

\begin{figure*}[t]
$\arraycolsep=1.0pt\def\arraystretch{0.6}
\begin{array}{ccc}
(a) \ \textrm{corotating} &  (b) \ \textrm{counter-rotating} 
& (c) \ \textrm{angle-integrated spectrum}
\\
{\centering\includegraphics[trim=0.0cm 1cm 1cm 1cm,clip=true,width=0.68\columnwidth]
{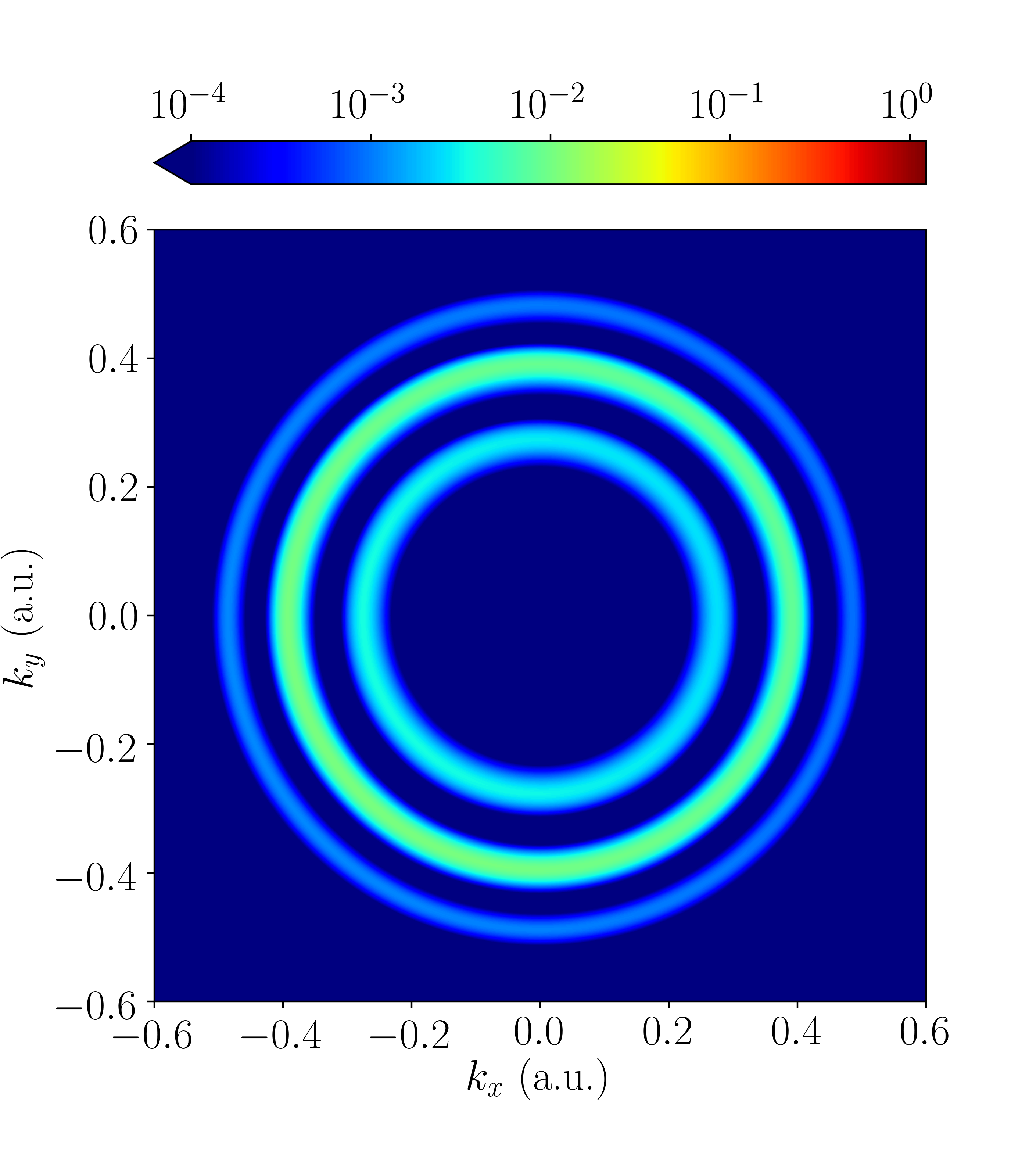}
}
&
{\centering\includegraphics[trim=0.0cm 1cm 1cm 1cm,clip=true,width=0.68\columnwidth]
{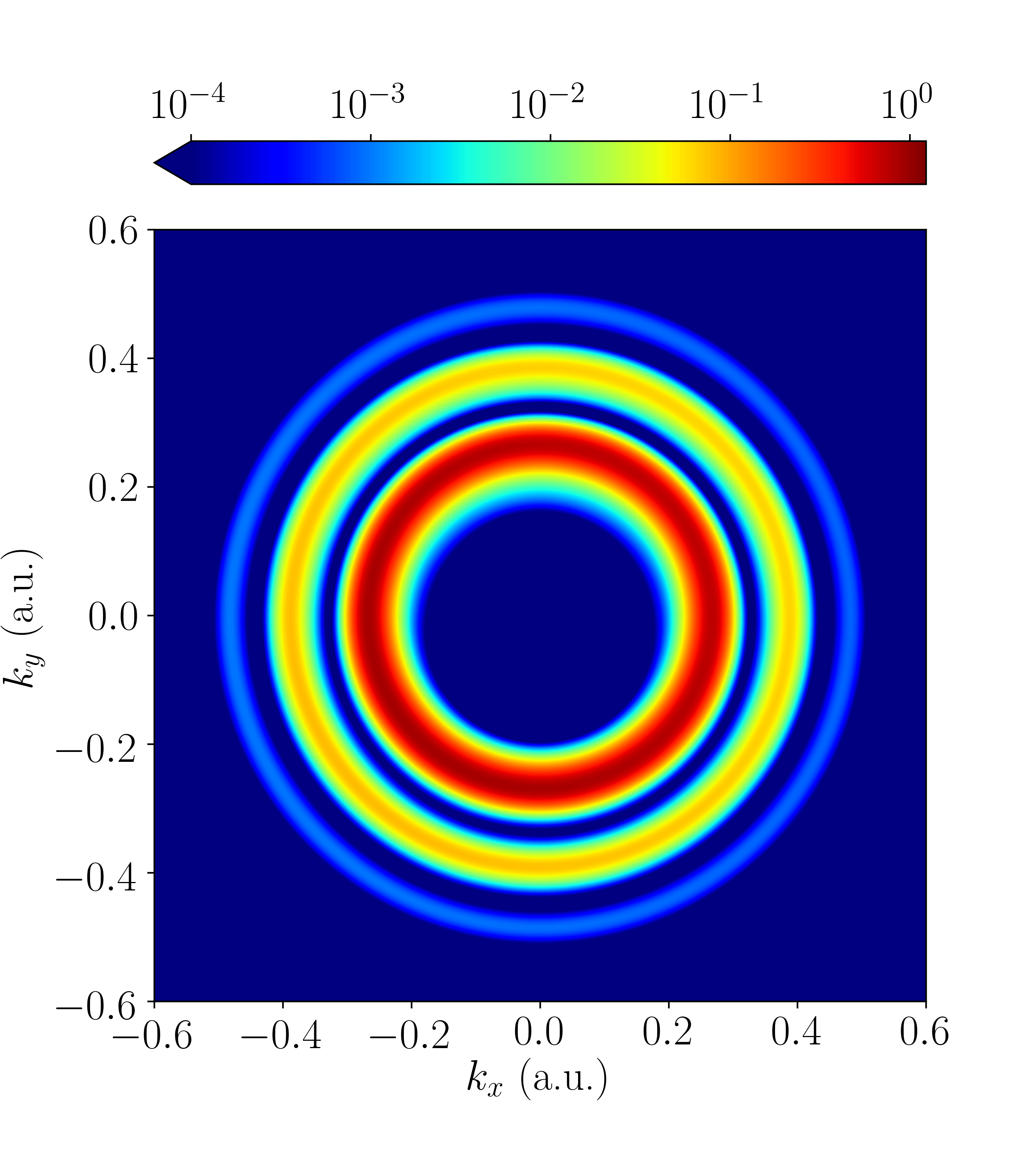}
}
&
\includegraphics[width=0.64\columnwidth]{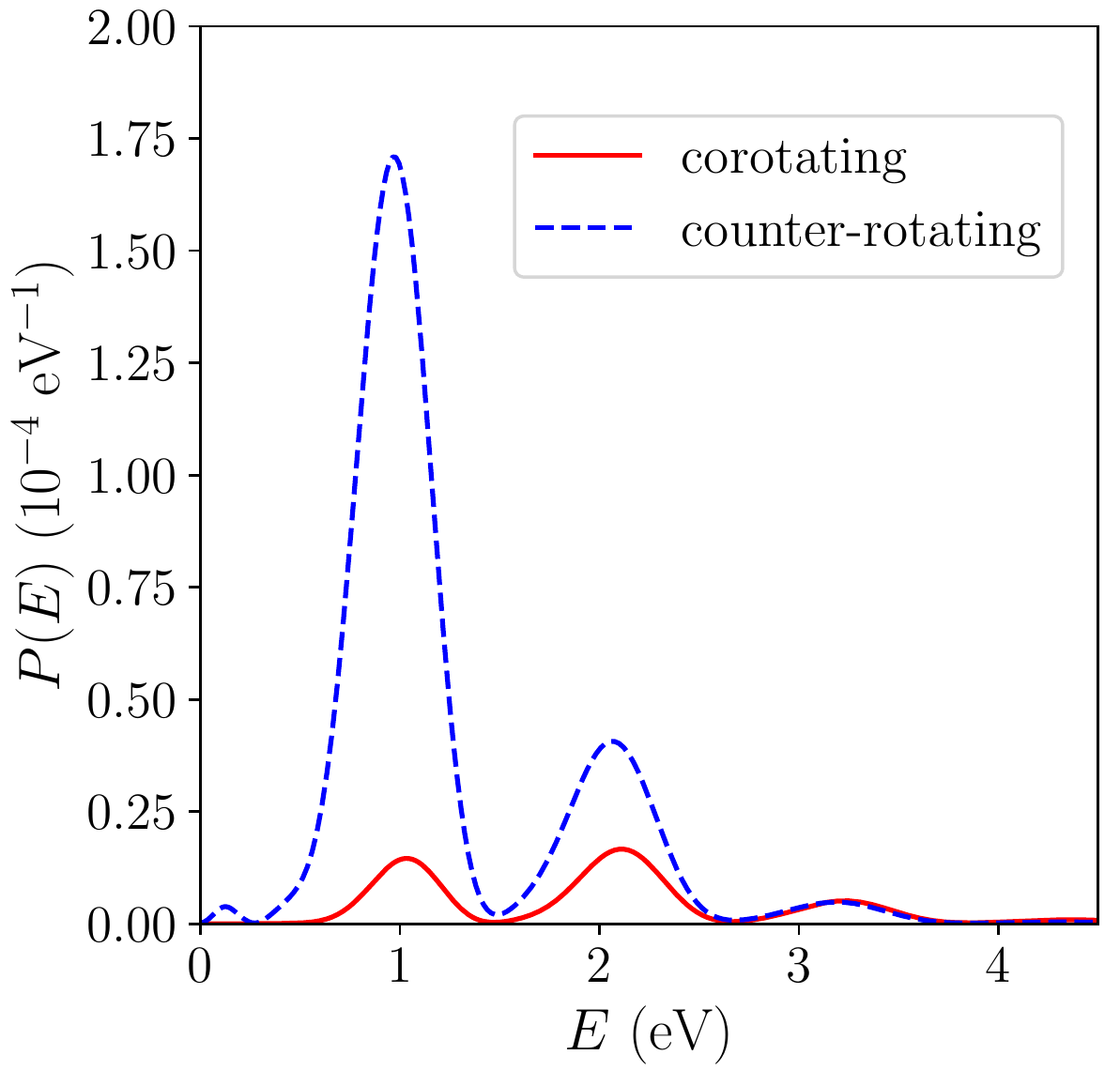}
\end{array}
$
\caption{(Color online) Photoelectron momentum distributions in the polarization plane for (a) corotating and (b) counter-rotating electron detachment, following four-photon detachment from F$^-$ by a right-hand circularly polarized laser pulse with a carrier wavelength of 1064 nm, a duration of $N_c = 6$ cycles, and a peak intensity of $I_0 = 3 \times 10^{12}$ Wcm$^{-2}$. The angle-integrated distributions (c) (integrated over azimuthal angle $\phi$) demonstrate the differing energy dependence of co- and counter-rotating electrons.} 
\label{4wfig}
\end{figure*}
%

\subsection{Four-photon detachment}

Figure \ref{4wfig} presents photoelectron momentum and energy distributions in the polarization plane, for four-photon detachment from F$^-$ by a 1064-nm, 6-cycle, right-hand circularly polarized pulse. The distribution now contains three distinct peaks, due to two significant above-threshold photon absorptions.
As in Fig.\;\ref{3wfig}, a strong energy-dependent electron rotational asymmetry is evident. Detached counter-rotating electrons dominate the signal close to threshold, before dying away relatively rapidly for higher-order ATD processes. However, in contrast to Fig.\;\ref{3wfig}, counter-rotating electrons detach more readily than corotating electrons after one above-threshold photon absorption. Near-parity between the detachment yields for co- and counter-rotating electrons is only established by the second above-threshold peak.

Aside from the disparity in their respective magnitudes, a noticeable difference in the energy-dependence of co- and counter-rotating electrons appears: co-rotating electrons make their strongest contribution at the first above-threshold peak, whereas the contribution from counter-rotating electrons decreases monotonically with excess energy. Here we observe the first indication of similarity with existing results using ARM and PPT theories for Kr \cite{barth2011,kaushal2015} and Ar \cite{eckart2018}, which have observed strong counter-rotating electron yields close to threshold, with the weaker signal of corotating electrons strengthening at higher energies. This trend is made clear in the energy distribution shown in Fig. \ref{4wfig}(c), where the counter-rotating electron contribution falls rapidly, while the corotating contribution peaks at around 2 eV.

\subsection{Five-photon detachment}
\begin{figure*}[t]
$\arraycolsep=1.0pt\def\arraystretch{0.6}
\begin{array}{ccc}
(a) \ \textrm{corotating} &  (b) \ \textrm{counter-rotating}
& (c) \ \textrm{angle-integrated spectrum}
\\
{\centering\includegraphics[trim=0.0cm 1cm 1cm 1cm,clip=true,width=0.68\columnwidth]
{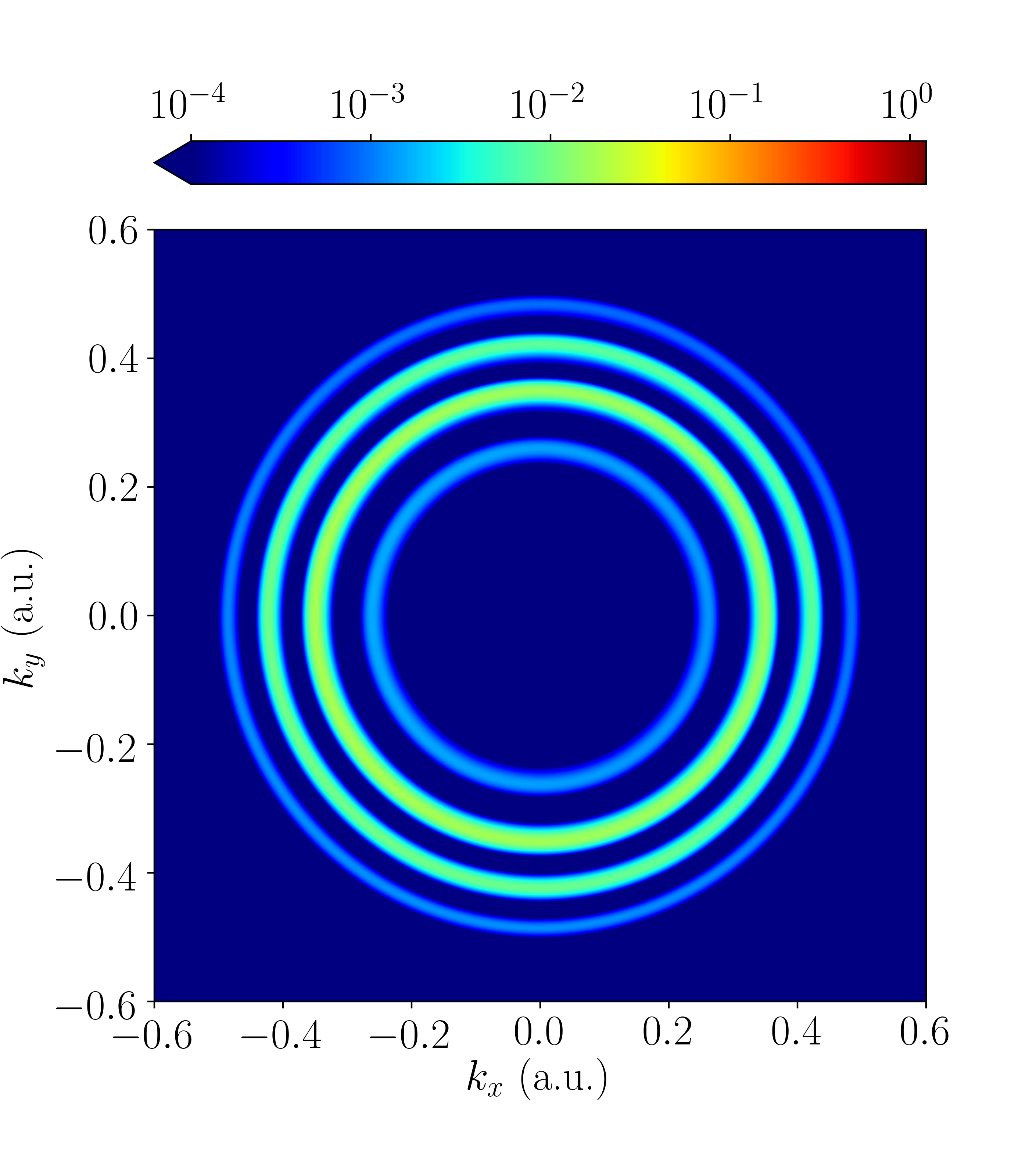}
}
&
{\centering\includegraphics[trim=0.0cm 1cm 1cm 1cm,clip=true,width=0.68\columnwidth]
{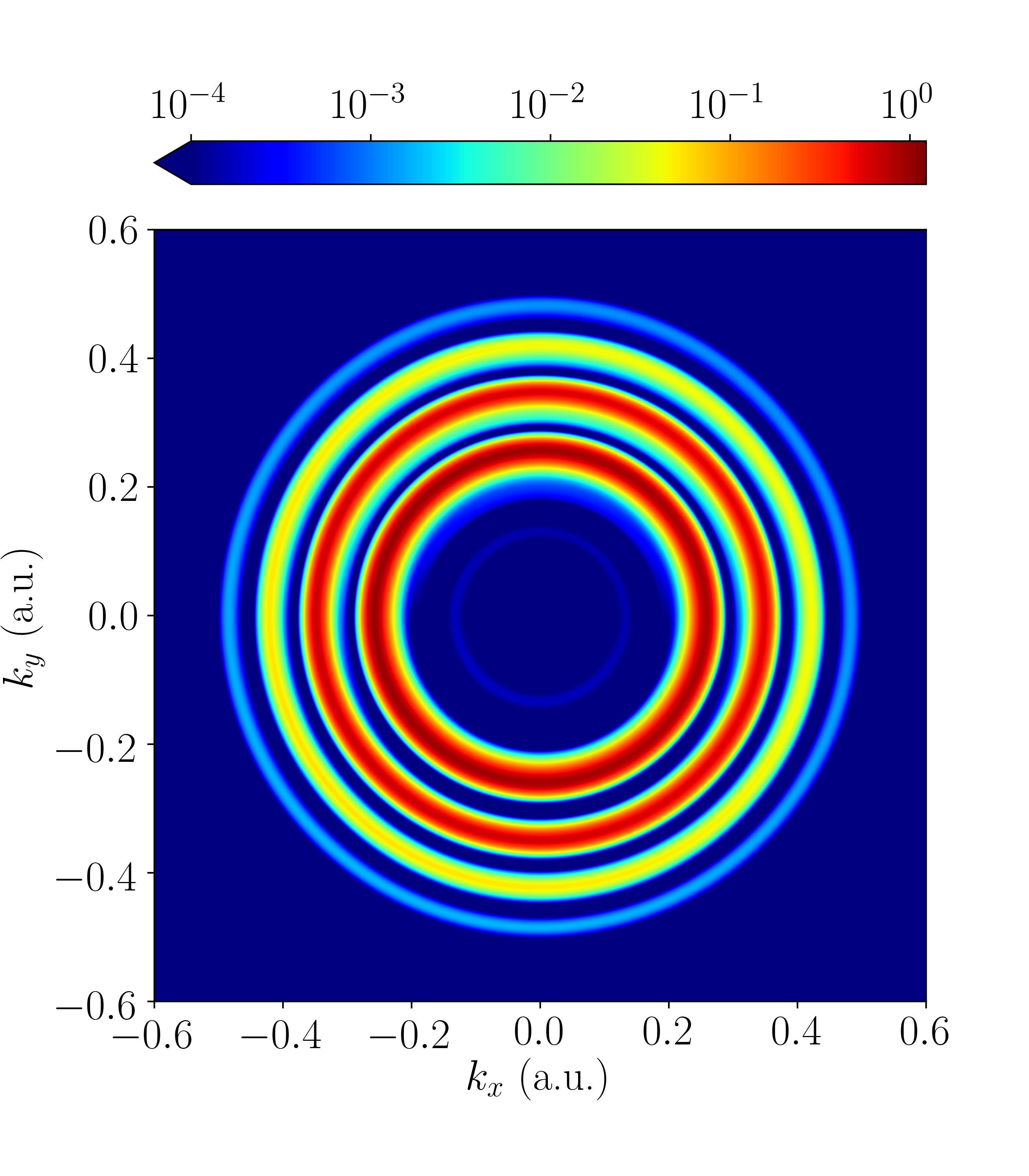}
}
&
\centering\includegraphics[width=0.64\columnwidth]{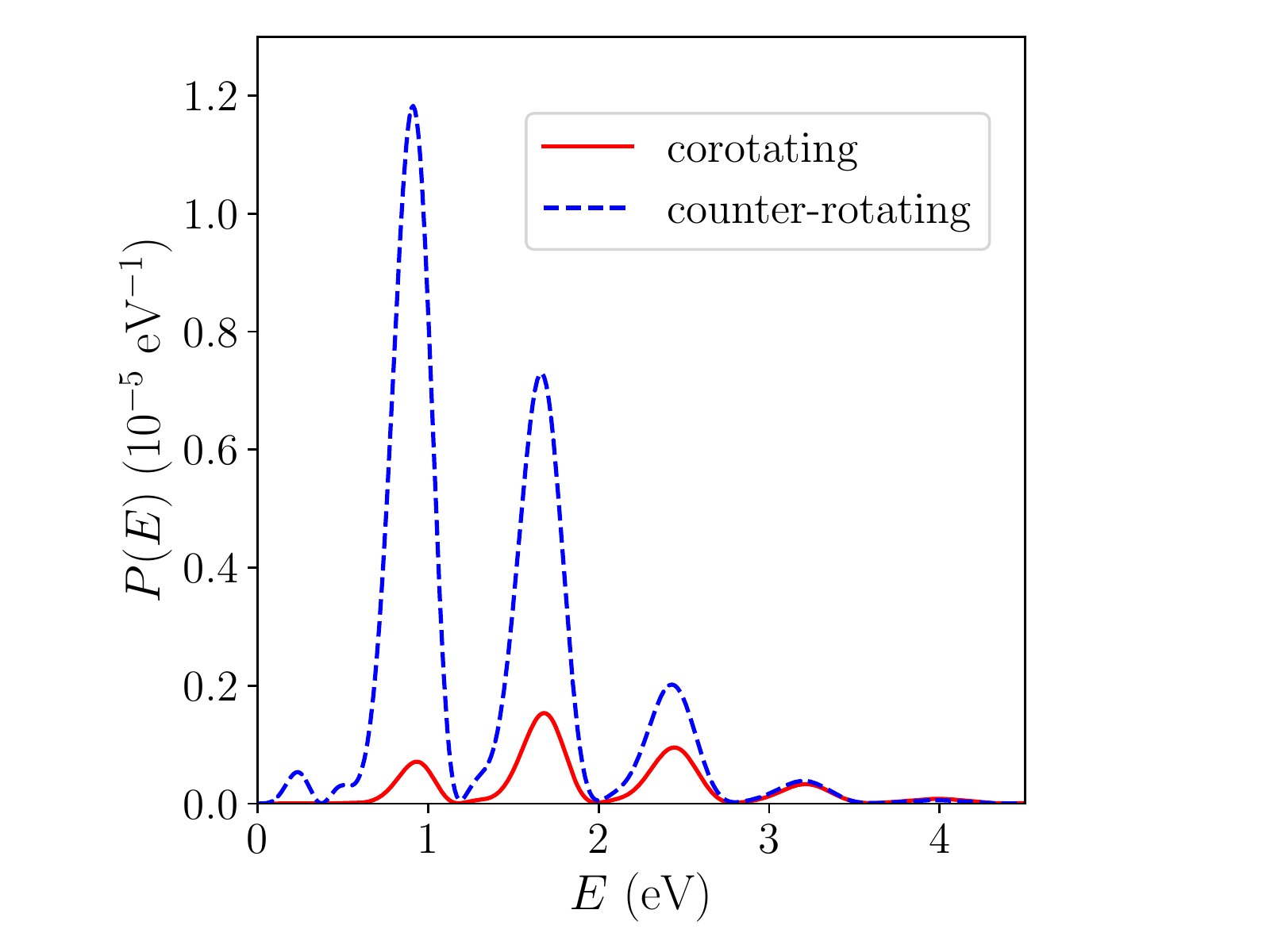}
\end{array}
$
\caption{(Color online) Photoelectron momentum distributions in the polarization plane for (a) corotating and (b) counter-rotating electron detachment, following five-photon detachment from F$^-$ by a right-hand circularly polarized laser pulse with a carrier wavelength of 1560 nm, a duration of $N_c = 8$ cycles, and a peak intensity of $I_0 = 2 \times 10^{12}$ Wcm$^{-2}$. The angle-integrated distributions (c) (integrated over azimuthal angle $\phi$) demonstrate the differing energy dependence of co- and counter-rotating electrons.} 
\label{5wfig}
\end{figure*}

Figure \ref{5wfig} presents photoelectron momentum distributions in the polarization plane, for five-photon detachment from F$^-$ by a 1560-nm, 8-cycle, {\mbox{$2\times10^{12}$-W/cm$^2$}}, right-hand circularly polarized pulse. Above-threshold detachments are considerably more prominent at this wavelength, particularly for counter-rotating electrons, where the two innermost peaks differ in magnitude by only around 25\%. Many of the features of Fig.\;\ref{4wfig} are visibly extended in the two cases shown in Fig.\;\ref{5wfig}, and the presence of multiple above-threshold peaks allows a trend to be discerned. Counter-rotating electrons are preferentially detached close to threshold, before their contribution monotonically decreases as excess energy increases. Corotating electrons again appear more likely to be detached in above-threshold processes than close to threshold. A strong rotational asymmetry is clear at a number of ATD peaks, with symmetry approached only at the outermost ATD peak.  These observations are again reminiscent of those found for noble-gas targets using PPT \cite{barth2011} and ARM \cite{kaushal2015} approaches.


\subsection{Comparison with analytical expressions}

In approaching the strong-field limit, it is interesting to determine if such an effect is predicted in the analytical PPT and ARM theories. By considering a monochromatic circularly-polarized pulse of frequency $\omega$, Eq.\;(19) in Ref.\;\cite{barth2011} and Eq.\;(B31) in Ref.\;\cite{kaushal2013} provide an analytical expression for the $n$-photon ionization rate, and its dependence on bound-electron $m_l$. When considering detachment of $p$ electrons from negative ions, the respective expressions in Refs.\;\cite{barth2011} and \cite{kaushal2013} become identical, giving the ionization rate  $w_n^{p_{m_l}}$ whose $m_l$ dependence is expressed simply by
\beq
w_n^{p_{m_l}} \propto \left[\sqrt{\frac{\zeta^2+\gamma^2}{1+\gamma^2}}-\zeta\; {\rm sgn}(m_l)\right]^2 .
\label{ppt}
\eeq
Here, $\zeta=(2n_0/n)-1$ is a dimensionless parameter dependent on the minimum number of photons for detachment, $n_0$, and the total number of photons absorbed, $n$. For a circularly-polarized field, $n_0$ is related to the ionization potential $I_p$ and ponderomotive energy $U_p$  by
\beq
n_0 = \frac{I_p + 2U_p}{\omega}.
\nn
\eeq
Finally, $\gamma=(I_p/2U_p)^{1/2}$ is the Keldysh parameter.

\begin{figure*}[t]
$\arraycolsep=1.0pt\def\arraystretch{0.6}
\begin{array}{cc}
{\centering\includegraphics[width=\columnwidth]{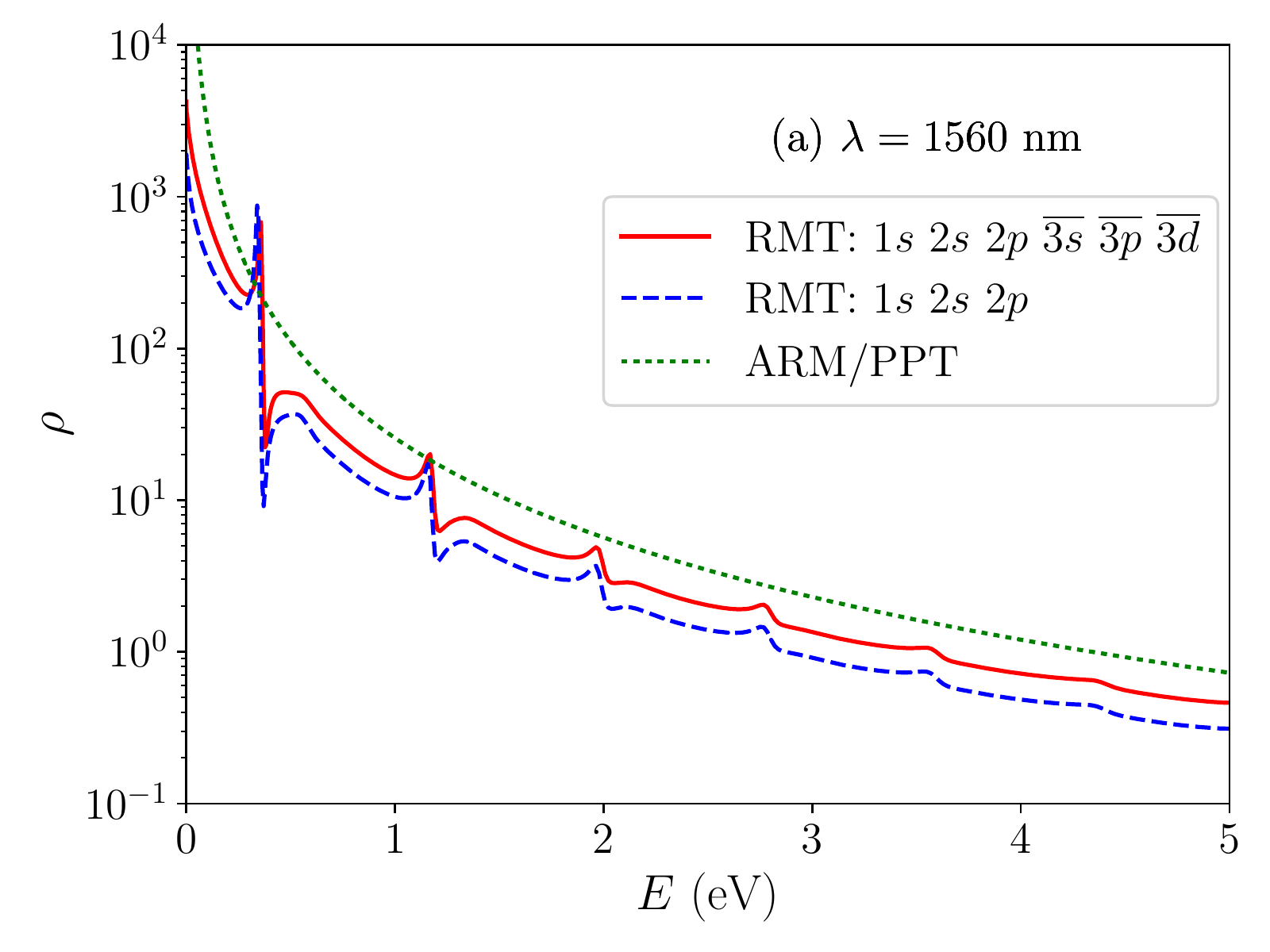}
}
&
{\centering\includegraphics[width=\columnwidth]{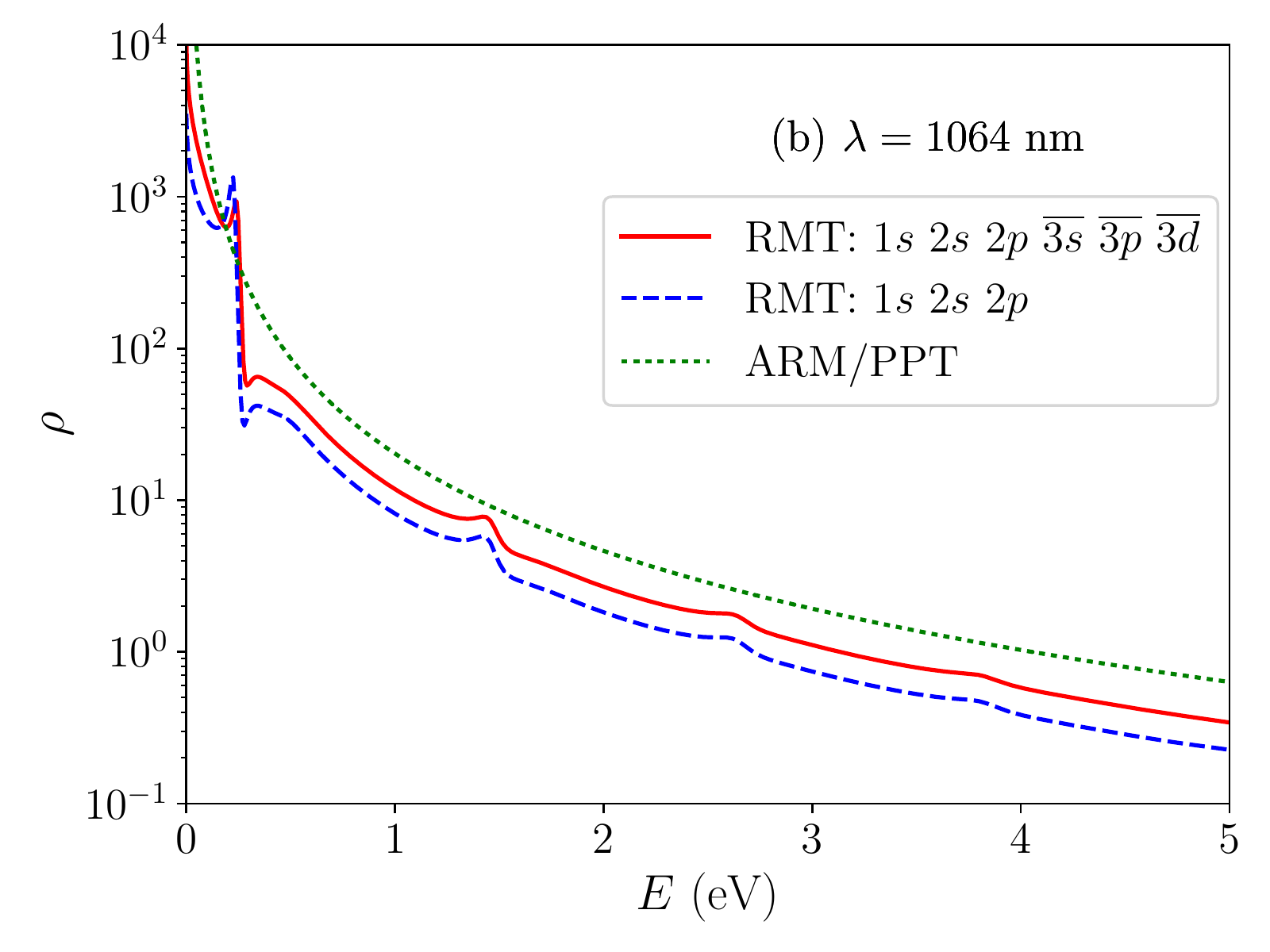}
}
\end{array}
$
\caption{(Color online) Energy-dependent ratio of $p_{-1}$ electrons to $p_1$ electrons following photodetachment from F$^-$, initiated by right-hand circularly polarized laser pulses of carrier wavelength (a) 1560 nm and (b) 1064 nm. The ratio given by ARM and PPT theories (Eq.\;\eqref{ppt}) is compared against two RMT calculations: one using an atomic structure that includes the Hartree-Fock $1s$, $2s$ and $2p$ orbitals, and another that includes the $\overline{3s}$, $\overline{3p}$ and $\overline{3d}$ pseudo-orbitals. The photoelectron energy is derived from the parameters of Eq.\;\eqref{ppt} as $E=(n-n_0)\omega$.} 
\label{ratios}
\end{figure*}

%


We note that the ARM approach has been formulated for both infinitely long (monochromatic) and finite laser pulses \cite{kaushal2015}. However, in the former case, ARM theory yields a simple, closed-form result for the $m_l$-dependent ionization rate, [Eq.\;\eqref{ppt}], which not only concurs with that of PPT theory, but which is also convenient and practical, being free of complex, numerical calculations. Since our purpose in this work is to gain insight into the general, energy-dependent trend of the asymmetry ratio, and to draw a qualitative comparison between methods, we shall employ only the ARM and PPT results for monochromatic fields.


At threshold (i.e. $n=n_0$), Eq.\;\eqref{ppt} predicts that only $p_{-1}$ electrons are detached. The contribution of $p_1$ electrons gradually rises as $n$ increases. When $\zeta=0$ (i.e. $n=2n_0$), the rate is independent of $m_l$, and hence $p_{\pm1}$ electrons are predicted to detach with equal probability. For {\mbox{$n\gg n_0$, $p_1$}} electrons dominate the detachment yield, although the yield at such energies is insignificant. Such variation has been observed in experiment \cite{eckart2018}, with ARM predictions agreeing well with the measured spectra and SAE calculations. Therefore, Eq.\;\eqref{ppt} predicts a photoelectron energy spectrum (or momentum distribution) whose first $n_0$ peaks are overwhelmingly due to detachment of a counter-rotating electron, whereafter corotating-electron detachment strengthens. Such behaviour is in fact prevalent in ionization of $p$ electrons from noble gases, and has been demonstrated in Ref.\;\cite{barth2011} for ionization of $4p$ electrons in Kr, and in subsequent studies for a variety of atomic targets exposed to few-cycle pulses \cite{kaushal2015,kaushal2018a,kaushal2018b}.

The relative contributions of counter-rotating and corotating electrons are commonly quantified using their energy-dependent ratio, $\rho$, which PPT and ARM theories provide in the convenient form
\begin{equation}
    \rho =
    \frac{w_n^{p_{-1}}}{w_n^{p_1}}
    =
    \left[
    \frac{
    \sqrt{\frac{\zeta^2+\gamma^2}{1+\gamma^2}}+\zeta
    }
    {
    \sqrt{\frac{\zeta^2+\gamma^2}{1+\gamma^2}}-\zeta
    }
    \right]^2.
    \label{pptrat}
\end{equation}
We stress that this result is based on two main assumptions --- a monochromatic laser pulse, and an initial state in which correlation effects are neglected.

In RMT, we calculate this quantity by integrating the photoelectron momentum distribution over the angular variables to obtain the energy spectrum, and taking the ratio of the respective spectra for co- and counter-rotating electrons.

Figure \ref{ratios} shows the ratio of counter-rotating to corotating electron detachment yields, as a function of the photoelectron energy, calculated using the RMT and ARM/PPT methods [Eq.\;\eqref{pptrat}] at a wavelength of (a) 1560 nm and (b) 1064 nm. Here we show two RMT calculations --- one that includes $\overline{3s}$, $\overline{3p}$ and $\overline{3d}$ pseudo-orbitals in the atomic structure description of F, and one in which Hartree-Fock $1s$, $2s$ and $2p$ orbitals are used (see Sec.\;\ref{calcp} for details).


We begin by comparing the RMT calculation including pseudo-orbitals with the ARM/PPT ratio. It is clear from Fig.\;\ref{ratios} that both methods predict a similar qualitative energy dependence at each wavelength, with counter-rotating electrons strongly dominating near threshold, before falling away at higher excess energies. The series of rapid resonance-like variations in the ratio calculated using RMT occur at energies between the photoelectron peaks (see Fig. \ref{5wfig}(c)), and are due to division of small yields. We do not attempt a comparison at 800 nm, due to the limited number of significant above-threshold detachment peaks in the distributions of Fig.\;\ref{3wfig}.

We emphasize that a high degree of quantitative agreement between the methods is not expected, given the differences in pulse lengths and the differing accounts of correlation in the initial state. These comparisons demonstrate the capture of the energy-dependent disparity between co- and counter-rotating electrons by an {\em ab initio} approach. It allows an initial assessment of the trends predicted by the ARM and PPT theories, that are commonly used for strong-field processes in circularly polarized laser pulses.

%


We now examine the degree to which the distributions are influenced by the atomic structure of the residual neutral. To do this, we perform calculations using the Hartree-Fock atomic structure model. We find that the total detachment yields calculated using this model are around 40\% higher than those obtained with the more complete model (results not shown), despite the artificial shift in the ground-state binding energy of F$^-$. This sensitivity to the atomic structure description is commensurate with that observed in previous calculations for linearly-polarized fields \cite{hassouneh2015}.

In both cases shown in Fig.\;\ref{ratios}, we find that the calculations using Hartree-Fock structure yield a ratio which is around 30\% lower than that obtained when pseudo-orbitals are included. This difference appears to be weakly energy-dependent, thus preserving the general trend of $\rho$. 
This implies that variations in the atomic structure description have a noticeable impact on the magnitude of $\rho$, but do not modify the general trend significantly.
Nonetheless, 
from Fig.\;\ref{ratios} we conclude that a simple Hartree-Fock model proves inadequate for the accurate treatment of processes in which short-range correlations influence the dynamics. This was previously demonstrated in the context of electron rescattering in our previous study of strong-field detachment in this system \cite{hassouneh2015}. The flexibility offered by the present RMT method, with respect to the degree of atomic structure retained in the calculations, thus enables a proper assessment of the role of electron-electron interactions in strong-field processes (often simply neglected in analytical models).


\section{Conclusions}
We have performed {\em ab initio}, nonperturbative calculations of photoelectron momentum distributions for F$^-$, in circularly-polarized laser pulses, using the RMT method. We have established the contributions from corotating and counter-rotating electrons, and demonstrated the well-known preference for strong-field detachment of counter-rotating electrons. Furthermore, we have observed a strong variation in this preference with photoelectron energy. Close to threshold, a strong rotational asymmetry favours detachment of counter-rotating electrons. As excess energy increases, the degree of rotational asymmetry gradually decreases, with equal partitioning only established at high excess energies, where detachment yields are negligible. Conveniently, the quantity of interest, the energy-dependent ratio of counter-rotating to corotating electrons, is provided analytically by the ARM and PPT approaches. We find good qualitative agreement between the predictions of these methods, and our numerically-calculated ratios. This demonstrates the ability of both numerical and analytical methods alike to capture the asymmetric distribution of detachment yields in circularly-polarized fields, and provides a valuable verification of their predictive power. Our work further underlines that the RMT method is capable of capturing the multiphoton dynamics of a truly multielectron atom in circularly-polarized laser light. We also highlight that RMT provides a means of inferring sensitivities to the atomic structure description, that are more difficult to gauge in other approaches.


\section*{Acknowledgments}
We acknowledge Jakub Benda and Zdenek Ma\v{s}\'{i}n for their collaboration in developing and maintaining the RMT code. The data presented in this article may be accessed at Ref.\;\cite{pure}. The RMT code is part of the UK-AMOR suite, and can be obtained for free at Ref.\;\cite{repo}. This work benefited from computational support by CoSeC, the Computational Science Centre for Research Communities, through CCPQ. DDAC acknowledges financial support from the UK Engineering and Physical Sciences Research Council (EPSRC). ACB, HWvdH and GSJA acknowledge funding from the EPSRC under grants EP/P022146/1, EP/P013953/1 and EP/R029342/1. This work relied on the ARCHER UK National Supercomputing Service (\href{www.archer.ac.uk}), for which access was obtained via the UK-AMOR consortium funded by EPSRC.

\appendix




\end{document}